\documentclass[a4paper,12pt]{article}

\usepackage{a4wide}
\usepackage{epsfig}
\usepackage{amssymb}
 \usepackage{amsthm}

\newcommand{\sverb}[1]{{\small \texttt{#1}}}
\theoremstyle{definition}
\newtheorem{example}{Example}

\begin{document}




\title{The role of concurrency in 
      an evolutionary view of programming abstractions}


\author{Silvia Crafa\\
Dipartimento di Matematica - Universit\`a di Padova\footnote{
This work was partially supported by the University of Padova under the PRAT project “BECOM”.}}
\date{}
\maketitle

\begin{abstract}
In this paper we examine how concurrency has been embodied in
mainstream programming languages. In particular, we rely on the
evolutionary talking borrowed from biology to discuss major historical
landmarks and crucial concepts that shaped the development of
programming languages. 
We examine the general development process, occasionally deepening
into some language, trying to uncover evolutionary lineages related to
specific programming traits. We mainly focus on concurrency,
discussing the different abstraction levels involved in present-day
concurrent programming and emphasizing the
fact that they correspond to different levels of explanation.
We then comment on the role of theoretical research on the quest
for suitable programming abstractions, recalling the importance of
changing the working framework and the way of looking every so often.
This paper is not meant to be a survey of modern mainstream
programming languages: it would be very incomplete in that sense. 
It aims instead at pointing out a number of remarks and connect
them under an evolutionary perspective, in order to grasp
a unifying, but not simplistic, view of the programming languages
development process.

\end{abstract}




\section{Introduction}
\label{sec:intro}

It is widely acknowledged that theoretical research and applications
development evolve at different speeds, driven by different
aims. Research theory investigates new problems and new ideas, taking
its time to study different solutions that tackle problems from
different points of view. It often happens that while deeply studying
one of these solutions, new ideas or entirely new problems emerge,
opening the way to entirely new (sub)theories.
The development of applications, instead, is much more oriented
towards effective solutions. Moreover, the choice between different
solutions is often dictated by technology constraints or market
constraints such as cost-effectiveness and rapid productivity.
The (relative) low speed and the wide scope of the theory often
result in a (relative) lack of integration of results. The
theoretical outcomes of research are so abundant that their
integration and assimilation is quite hard. Nevertheless,
incorporating different results under a unifying or coherent view,
both formally and epistemologically, would be of great help for the
progress of the knowledge. 
On the other hand, the speedy development of application solutions
tends to miss the opportunities offered by theoretical results that
have been already established but not yet fully applied.

While the percolation of theoretical results into applications
intrinsically requires some time, it is helpful to remind taking a
look from theory towards applications and from applications to theory
every so often. 

This paper is then written in this spirit. We examine the history of
popular programming languages to expose how concurrency has been
incorporated into the mainstream programming. This excursus is not
meant to be a survey of concurrent languages; it would be very
incomplete in that sense. We aim instead at identifying a number of
historical landmarks and crucial concepts that shaped the development
of programming languages. Moreover, rather than fully describing these
major points, we focus on their connection under an evolutionary
perspective, using the biological theory of evolution as an
instructive metaphor. We then borrow the evolutionary
talking from biology, using it as an explanatory tool 
to more deeply reflect on some concepts and to stimulate the
development of meta-knowledge about the history of programming
languages. 

\paragraph{Structure of the paper}
In Section 2 we start our excursus of the history of popular
programming languages from an evolutionary perspective. We first
examine the general development process, and we then deep into
some language, trying to uncover evolutionary lineages
related to specific programming traits. In Section 3 we consider
concurrency abstractions: we put forward three different concurrency
models used in mainstream programming, emphasizing the fact that they
correspond to three different levels of explanation. The evolutionary
excursus is then completed in Section 4 with a discussion of the
impact of the Clouds and Big Data technologies in programming
languages. In Section 5 we discuss the role of the theoretical
research in the evolutionary scenario, putting forward its ability
of promoting and testing language mutations. We conclude in Section 6
with final comments.

\section{An evolutionary look at mainstream programming}
\label{sec:quest}

The research about programming languages can be described in many
different ways. An interesting approach is looking at languages in a
timeline perspective, trying to grasp the evolutionary process that
guided (or that unfolded behind)
 the fortune of mainstream programming languages.  
It has been suggested (e.g.,\cite{EvoEco,PunctPL}) that information
technology innovations occur mainly through the combination of
previous technologies. Even living structures are the result of a
widespread reuse and combination of available elements, but in biology
established solutions are seldom replaced, while the introduction of
new simple technological elements can completely reset the path of
future technologies.
Another key feature of technological evolution is that it is mainly
guided by planned designs, that have no equivalent with natural
evolution: technology designers seek optimality criteria in terms of
correctness, efficiency, cost and speed, and they outline new goals
and expectations. Nevertheless, long-term trends and the
diversification effects of contingencies, also due to
social and economical factors, can only be captured a posteriori.

We then give here a very incomplete and extremely
partial outline of what could be called the modern history of
mainstream programming languages. If we try the exercise of listing
major languages in a total, time-based ordering, we observe that even
a very rough ordering requires to choose a non-trivial criterion.
Should we list languages according to when they have been
invented or according to when they became popular? Interestingly, such
a question exposes the gap between why a language has been
invented and why it became popular. For instance Objective-C was
designed long before the advent of mobile devices but its popularity
greatly depends on the boost in the proliferation of apps for Apple
mobile tools (see the Objective-C's TIOBE Index in \cite{ObectiveC-Tiobe}).

\begin{figure}[th]
\begin{center}
\includegraphics[width=12cm]{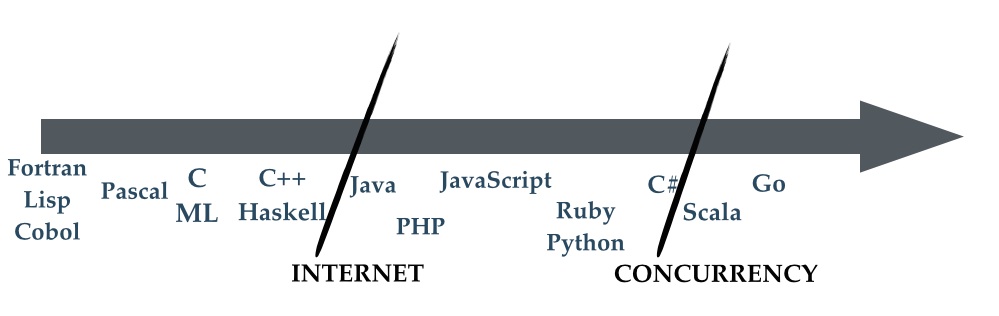}
\caption{Timeline}
\label{fig:timelineP}
\end{center}
\end{figure}

A fair solution to the ordering problem is then not to linearize the
languages, but the major evolutionary leaps that marked the area, as
depicted in Figure~\ref{fig:timelineP}.

With the introduction of languages such as Fortran, Lisp, Cobol and
Algol around 1950s-1960s we can start talking about modern languages,
that rely on primitives such as \sverb{if}, \sverb{goto},
\sverb{continue} to add structure to the code. Around 1970s-1980s the
advent of languages like C, Simula, Smalltalk, ML and Prolog marks the
rise of modern programming \emph{paradigms}, such as imperative
programming, functional programming, logic programming and
object-orientation. 
The case of object-oriented programming (OOP) is particularly
instructive. In its well known paper 
``The free lunch is over''~\cite{FreeLunch05}, Herb Sutter recalls
that even if OOP dates back in 1960s, object-orientation didn't
become the dominant paradigm until the 1990s. He also observes that
the shift appeared 
\begin{quote}
``when the industry was driven by requirements to
write larger and larger systems and OOP's strengths in abstraction and
dependency management \emph{made it a necessity} for
achieving large-scale software development that is economical,
reliable, and repeatable''. 
\end{quote}
Correspondingly, the killer application for OOP and its modularity,
encapsulation and code reuse principles, has been recognized to be the
development of GUIs (graphical user interfaces). 
The key observation here is that the object-oriented model, well
studied in academia, became pervasive in applications and mainstream
programming only when driven by critical industry requirements.

Similar evolutionary leaps in the history of programming
languages can be traced also in recent years. More precisely, we can
identify a number of {\it catalysts} that powered significant changes
in mainstream programming. First of all, the advent of the Internet
(and its appeal to the market) shifted the programming language goals
from efficiency to portability and security. This is the
scenario where Java came into the arena, and the JVM brought to the
fore the concept of virtual machine bytecode, already used, e.g., in
ML and Smalltalk. 
But the growth of the Web had an impact also on the popularity of
other languages: the so-called scripting languages, such as PHP,
JavaScript, Python, Ruby, which are well suited to 
the development of short programs to be embedded into web pages and
web servers.
Notice however that, besides web technologies, these languages 
are often the favorite choice of many general-purpose
programmers because of their high-level, declarative programming
model which enhances productivity and fast prototyping. It is worth
remarking here that the success of these dynamic languages as
general-purpose languages also comes as a reaction to the heavy and
verbose type discipline imposed by strongly typed languages such as
Java or C\#.

Another landmark in the history of programming languages, which
is directly related to the topic of this paper, is the popularity of
Concurrent Programming, whose catalyst is the fact that new,
efficient hardware can only be somehow parallel. Moore's law,
establishing that CPU performance doubles approximately every two
years, is still valid only because performance gains can nowadays be
achieved in fundamentally different ways: by means of CPU
hyperthreading (i.e. many threads in a single CPU) and multicore 
(many CPUs on one chip). However, in order to benefit from such a new
hardware, applications must be concurrent.
For the sake of clarity, we observe that parallel and concurrent
computing are different concepts, even if they are often used (also in
this article) as synonyms. They both refer to computations where more
than one thread of control can make progress at the same time. Parallel
computing stresses on the fact that many computations are actually
carried out simultaneously by means of parallel hardware, such as
multi-core computers or clusters. Concurrent computation generally
refers to tasks that may be executed in parallel either effectively,
on a parallel hardware, or virtually, by interleaving the execution
steps of each thread on a sequential hardware. Moreover, in the
context of programming languages, parallel tasks are generally sets of
independent activities that are simultaneously active, while
concurrent programs focus on the coordination of the interactions and
communications between different tasks.

As in the case of OOP, concurrent programming has been known since
1960s (e.g., it is a core aspect of any operating system), but
it became widespread and mainstream essentially only because of the
inexorability of parallel hardware. 
The greatest cost of concurrency, that also limited its accessibility,
is that (correct) concurrent programming is really hard and
refactoring sequential code to add concurrency is even
harder. Concurrency is so hard partly because of 
intrinsic reasons, such as dealing with nondeterminism, 
but also because of accidental reasons, like improper programming
models.  

\begin{figure}[th]
\begin{center}
\includegraphics[width=12cm]{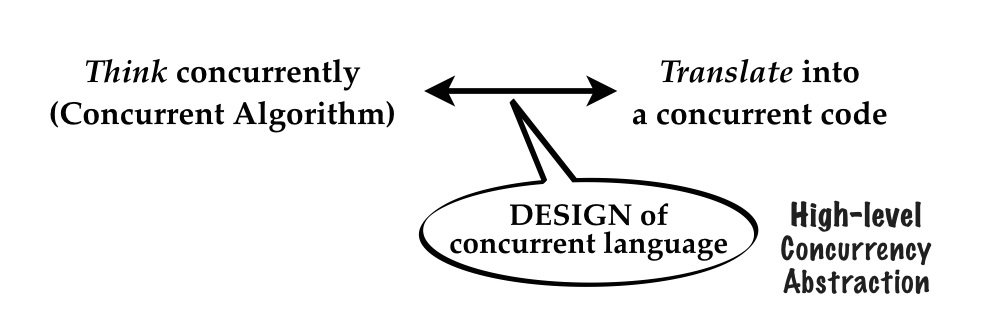}
\caption{Concurrency Gap}
\label{fig:ConcGap}
\end{center}
\end{figure}

As illustrated in Figure~\ref{fig:ConcGap}, concurrent programming
involves two phases. First the design of a concurrent algorithm or a
concurrent software architecture,
then its translation into a concurrent code. The feasibility
of such a translation clearly depends on the design choices of the
target programming language. Many concurrency models have been
studied, offering a range of solutions at very different abstraction
levels. Single Instruction Multiple Data (SIMD) architectures such as
General Purpose-GPUs are tailored to specific parallel hardware
devices, and efficient lock-free programming techniques require a
precise account of the processor's memory model. At the other extreme
stays the Actor model, which fosters declarative programming and
integrates well with OOP. Different concurrency models 
lead to different concurrent languages, e.g., Java, MPI, Erlang, Cuda,
that entail very different programming styles, each one with merits
and shortcomings. In particular, an 
effective high-level concurrent programming model is still
lacking. Such a deficiency can be rephrased in other terms: the quest
for satisfactory, high-level concurrency abstractions is still
open. 

We will further examine this point in Section~\ref{sec:concAbs}, but  
before diving into concurrency, we now make our look at mainstream
programming more concrete by applying the evolutionary explanation to
specific language traits.

\subsection{The quest for suitable programming abstractions}

After having discussed how large-scale industrial software, the
Internet and the popularity of parallel hardware had a pivotal role in
the development of mainstream programming, we now shift the point of
view and we move within some language, trying to uncover specific
evolutionary processes. 
Rather than looking at the history of a single language, 
we focus on how some key programming features evolved in time, in the
same spirit of the biological study of the evolution of specific
traits, like the wings, the eye or the thumb, across different
species.

In the realm of programming languages, a suitable definition of
heritable traits to be used for studying a sort of programming
language 
phylogeny is far from being clear. We then consider here the general
notion of \emph{programming abstractions} as the semantic traits 
that we can trace back in the history of programming languages. In
particular, we discuss some object-oriented abstraction, the
type discipline and the integration of functional programming with
other paradigms.

\paragraph{Object Oriented Languages} If we consider the object model
used by OOLs, we can
identify a tight ``evolutionary lineage'' starting from C++, passing
through Java and ending in Scala. For instance, the enforcing of
encapsulation has progressively increased along this lineage: C++'s 
friend functions violating encapsulation disappeared in Java, and
Scala adopted the Uniform Access Principle (which dates back to the
Eiffel language), that is, object's fields and 
methods are accessed through the same notation, preventing the
disclosure of any implementation details. 
Moreover, research and practice about
multiple inheritance conducted from C++'s superclass diamonds with
virtual inheritance, through Java's single inheritance and multiple
interfaces, up to Scala's mixins. 

This lineage also illustrates another trend in mainstream,
general-purpose, languages: the programming style becomes more
declarative and high-level, while implementation and efficiency
issues are progressively moved under the hood by increasing the 
complexity of the runtime. A simple example is
memory management: Java's automatic 
garbage collector takes fine-grained memory handling away from the
programmer's control. Additionally, Java's distinction between the
primitive type \sverb{int} and the class type \sverb{Integer}
disappeared in Scala. Primitive types exist in Java because they
allow for an efficient memory implementation that avoids the overhead
of class representation; Scala's solution (inherited from Smalltalk)
keeps a uniform object-oriented model in the language, that
only has the \sverb{Int} 
class type, and delegates to the compiler (and the JVM) the
implementation of the \sverb{Int} class type as Java's efficient
\sverb{int} type.

\paragraph{Typed Languages} The rise and fall of types in mainstream
programming is well represented in another evolutionary lineage of
languages: the path from C++, to Java, to Python/JavaScript, up to
Scala. In this case the object-oriented type system of C++ (among
other aspects) proved to be a valid support for early finding
``message-not-understood'' errors typical of OOP. Then Java pushed
ahead types to a strong typing discipline, at the cost of becoming
verbose and possibly cumbersome. As a reaction, untyped\footnote{This
  class of languages should be better called dynamically typed
  languages since type information is carried, and checked, at runtime
  (an interesting discussion about this topic can be found in the
  TYPES e-mail forum~\cite{TypesThread}). However, the 
  point here is that in these languages the source code, the one
  written by the programmer, is essentially untyped.} languages like
Python or JavaScript attracted programmers more interested in concise,
easier to read and faster to write programming. 
Ancient philosophers teach that ``virtue stands in the middle'',
indeed recent solutions are a compromise between the two extremes:
statically typed languages such
as Scala or Apple's Swift~\cite{Swift} dramatically reduced type
verbosity by improving the compiler's ability to infer
types\footnote{Type inference has a long and rich history,
  see~\cite{FWbook}: ``the standard reference is Hindley
  in 1969, though Hindley remarks that the results were known to Curry
  in the 1950s. Morris in 1968 also proposed type inference, but the
  widespread use of type inference did not happen until Milner's 1978
  paper.''}
. At the 
same time, statically typed versions of untyped languages have been
proposed, e.g., Microsoft's TypeScript~\cite{TypeScript} for
JavaScript and Mypy~\cite{Mypy,TypedPy} for Python.
%

\paragraph{Functional Programming} Another interesting example to be
looked at in an evolutionary perspective is the case of functional
programming. For a long time programming by means of immutable data and
higher-order functions had been confined to languages that have never
become mainstream. Pure functional programming has its merits, but
sometimes the imperative style is more natural and much easier to
reason about. C\# is possibly the first mainstream language that
clearly marks the integration of OOP and FP into a multiparadigm
language as a design goal. However, the full integration of these two
paradigms is better achieved in Scala, whose default variables are
immutable and a function is nothing else than an instance,
i.e. an object, of the class (implementing the trait)
\sverb{Function}. It is important to remark here that Scala borrowed
many of its features from previous languages, from Smalltalk's Uniform
Object Model, to OCaml, OHaskell and PLT-Scheme design choices. 
We just intend to highlight here which abstractions
have been recently brought to the fore as a result of an instance of
the recombination process mentioned at the beginning of this
section. To conclude, it is mandatory to 
observe that the recent standards C++11 and Java8 both extended the
language with lambda-expressions, that is 
with higher-order functions. Besides encouraging a more declarative
programming style, as illustrated in the example below, functional
programming has been proved to leverage parallel programming over data
structures, which brings us back to our main topic.

\begin{example}[The evolution of iterations]
Consider a list of people from which we want to find the age of the
oldest male. Let's focus on the Java language, even if this is not
restrictive. In what we could call the original Java style, we would
write something like the following code:
{\small
\begin{verbatim}
Person[] people = ...
int maxAge=-1;
for(int i=0; i<people.length; i++)
  if(people[i].getGender()==MALE && people[i].getAge()>maxAge)
     maxAge=people[i].getAge();
\end{verbatim}
}
\noindent
This iteration corresponds to a pure imperative programming style.
A pure object-oriented style would rather use an \sverb{Iterator} over
a \sverb{Collection}, rephrasing the pattern as follows:
{\small
\begin{verbatim}
Collection<Person> people = ...
Iterator<Person> it=people.iterator();
int maxAge=-1;
while(it.hasNext()){
  Person p=it.next();
  if(p.getGender()==MALE && p.getAge()>maxAge)
     maxAge=p.getAge();
}
\end{verbatim}
}
\noindent
Since Java5 we can rephrase the iteration in a more abstract style,
using the foreach construct:
{\small
\begin{verbatim}
Collection<Person> people = ...
int maxAge=-1;
for(Person p : people)
  if(p.getGender()==MALE && p.getAge()>maxAge)
     maxAge=p.getAge();
\end{verbatim}
}
\noindent
This style is more abstract in the sense that the iteration
variable is not just an index (or an iterator, which is a sort of
pointer), but the current element under examination. We have also
abstracted away iteration details like the 
size of the collection and the iteration increment step.
With Java8 iteration can be even more abstract, focusing on
\emph{what} we want to do, without going into \emph{how} to do it,
thanks to higher-order methods:
{\small
\begin{verbatim}
Collection<Person> people = ...
int maxAge= people.stream().filter(p -> p.getGender()==MALE)
                           .mapToInt(p -> p.getAge())
                           .max();
\end{verbatim}
}
\noindent
In this code the data structure (\sverb{Collection}) is used to
produce a stream of elements that proceed through a pipeline of
aggregate operations. Also notice that these operations correspond to
the \emph{map-reduce} functional programing pattern, where the
\sverb{mapToInt} method is the \emph{transformer} and \sverb{max} is
the \emph{combiner}. 
This new iteration style is a real advantage when the
code heavily handles large data structures, since it
is easily parallelizable by relying on \sverb{parallelstream}s rather
than sequential \sverb{stream}s. Indeed, in presence of parallel
streams the Java runtime partitions the stream into multiple
substreams and let the aggregate operations to process these
substreams in parallel by minimizing the synchronizations required by
the concurrent computation.
Anyway, it is interesting to observe such a drift toward a more
declarative, high-level, programming style.
\end{example}

\section{Concurrent programming and concurrency abstractions} 
\label{sec:concAbs}

After having looked at how programming abstractions evolved in other
areas, in this section we go back to concurrency.
As we said above, a number of different concurrency models have been
proposed. We recall here three models mainstream programming has been
attracted by: the Shared Memory model, the Message
Passing model, and the GP-GPU Concurrency model.
However, before discussing the three models, we emphasize the fact
that they correspond to three different \emph{levels of explanation}
(with reference to~\cite{Floridi}). An epistemological account of this
aspect is clearly out of scope, we just observe that tackling
concurrent programming and finding suitable primitives is much harder
than, say, functional programming, since concurrency affects many
levels: the hardware, the operating system, the language runtime, the
language syntax, and the logical level of algorithms. Here a quote
from Robin Milner's Turing lecture~\cite{MilnerTuring} is particularly
fitting: 
\begin{quote}
``I reject the idea that there can be a unique conceptual
model, or one preferred formalism, for all aspects of something as
large as concurrent computation [...] We need many levels of
explanation: many different languages, calculi, and theories for the
different specialisms'', and also ``Many levels of explanations are
indispensable. Indeed the entities at a higher level will certainly be
of greater variety than those lower down''.  
\end{quote}

\begin{figure}[th]
\begin{center}
\vspace*{-0.5cm}\includegraphics[width=10cm]{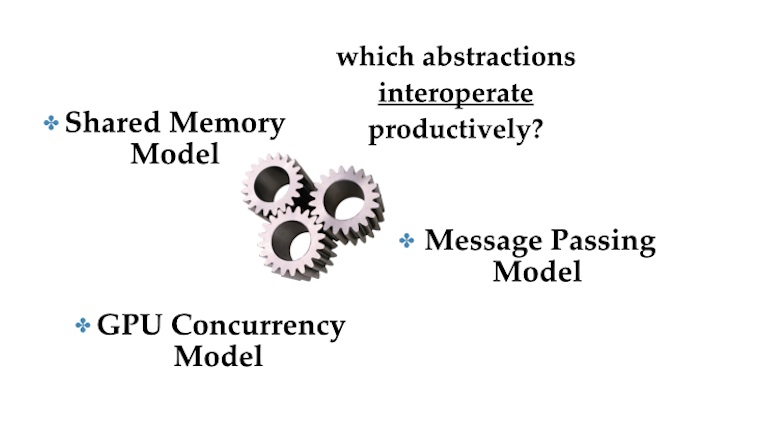}
\vspace*{-0.7cm}
\caption{The quest for convenient concurrency abstractions}
\label{fig:InterAb}
\end{center}
\end{figure}

\paragraph{Shared Memory Model}
In programming language terms, the Shared Memory model is
represented by Java-like threads: dynamically activated concurrent
flows of computation that communicate by means of synchronization on a
shared state. This model, relying on mutable shared state, is
well-suited to imperative programming. Moreover, it is a very natural
model for \emph{data-centric}, centralized algorithms and centralized
systems where different software components operate on shared data. 
However, protecting the shared state with
appropriate synchronization mechanisms has been shown to be very
difficult. Java-like locks and conditions (that are also available in
C\# and C++11) are error-prone and unscalable. Errors like data races,
deadlocks, priority inversion can be very subtle, and extensive
testing is difficult because of nondeterminism and the fact that locks
are not compositional. 

\paragraph{Message Passing model} 
When the application has a distributed nature and is
\emph{communication-centric}, shared-state is extremely
error-prone. In this case the Message Passing model is much more
valuable: the state is no longer shared but it is dispatched as a
message. Therefore, data races are avoided by construction and
deadlocks are infrequent. Message passing typically allows for a more
declarative programming style and more easily scales to a distributed
system, possibly at the cost of high communication and
coordination overheads. Notice that this model endorses immutable data
and programming techniques that use effects locally, selectively or
minimally, that is, functional programming. But there are not just
functional languages that provide support for message passing: for
instance the MPI standard is implemented in languages such as Fortran,
C and Java. The C-based Google's language Go 
offers message passing in a channel-based style. Anyway, whenever the
underlying paradigm is imperative, mutable variables are not immune to
subtle race conditions.

An interesting integration of the message passing model
with the object-oriented paradigm is represented by the Actor
model~\cite{Hewitt73,ActorsFirst}. An actor, like an object, has an
identity and 
reacts to messages in a single-threaded way in the same spirit of
method invocations. Moreover, encapsulation, modularization, the clear
distinction between interface and implementation are key notions also
in the design of actor systems. Interestingly enough, the following
quote from A. Key talking about OOL in 1998 fully applies also to 
actors: 
\begin{quote}
``the key in making great and growable systems is much
  more to design how its modules communicate rather than what their
  internal properties and behaviours should be'' \cite{AlanKay98}. 
\end{quote}
Notice that this is not an accident since the Actor model
has been proposed in the same years when 
languages with concurrent objects were studied (e.g.\cite{AghaT04}).
Today, the actor model is distinctive of Erlang, a purely functional
language with concurrency primitives that has been developed in the
1990s but whose popularity has grown in the 2000s due to the demand
for concurrent services. Nowadays Erlang is used in a number of
significant industrial projects such as Facebook Chat, GitHub,
WhatsApp. The actor model is also one of the
successful features of Scala, which is adopted for instance by
Coursera, LinkedIn, Ebay, essentially because it provides a type safe
language with scalable concurrency primitives on top of the JVM, which
is a mature technology platform. 
In is interesting to observe that, differently from Erlang, in the
case of Scala the original actor-based constructs are not primitive in
the language, but they are defined as a regular API whose
implementation emerges as an effective example, in the spirit of the
quest depicted in Section~\ref{sec:quest}, of how Scala's
object-oriented and functional abstractions productively
interoperate \cite{ScalaActorsImpl}\footnote{Recent 
  Scala releases deprecated the original Actor API in favor of the
  more efficient implementation offered by Akka Actors~\cite{Akka}}.

Another approach that is getting
increasing attention in the realm of communication-centric
computing is \emph{protocol-based computing}. In this view,
programming a concurrent (and distributed) system entails the design
of a precise communication protocol involving a, possibly dynamic, set
of interacting parties. Using the terminology of service-oriented
programming, if a service is represented as a sequence, or
more generally as a graph, of messages to be sent/received, then an
application can be viewed as service orchestration. However, web
service technology and languages (e.g., Jolie~\cite{Jolie}) are just
an example of what can be called protocol-centric
programming. Another example is offered by the Scribble
language~\cite{Scribble} and the other outcomes of the rich theory of
behavioural types and session types that are percolating at the level
of programming languages. 

\paragraph{General Purpose-GPU programming}

Over the years the Graphics Processing Unit (GPU) and its
massively parallel architecture proved to be convenient also for
general purpose usage, particularly for some class of algorithms and
applications where large amount of data need to be parallely
processed. Physics simulations, ray tracing, bioinformatics,
evolutionary computation, machine learning, oil exploration,
scientific image processing, linear algebra, statistics, 3D
reconstruction, bitcoins mining and even stock options pricing, and
many more fields and disciplines can benefit from this architecture.  
GPUs use the ``single instruction, multiple data''(SIMD) architecture,
in which multiple processors execute the same instructions on
different pieces of data. This allows the control flow
computation to be shared amongst processors, so that 
more of the hardware is devoted to instruction execution. 
However, given this architecture, 
GP-GPU programming requires a specific algorithmic
thinking and a corresponding programming model. In 2006 NVIDIA
launched CUDA, a parallel computing platform on  GPUs
comprising a programming model and an extension to the C programming
language. However, in the CUDA programming model it is fundamental to
precisely know the underlying architecture and its consequences on
the performance of programs. It is essential to be aware of how (sets
of) threads are mapped onto the hardware, how they are executed and
scheduled, and to know the peculiarities of memory access patterns.
Therefore CUDA brings about an extremely low-level programming, which
allows C/C++ programmers to significantly fine-tune the applications
performance, at the price of sacrificing high-level abstraction. 

The recent explosion of interest in  GP-GPU, both from the
research community and the industry is changing the situation. In
particular, efforts are made 
to unify host and device programming. On the hardware side, the
memories of the CPU and GPU are physically distinct but CUDA 6
very recently (\cite{CUDA6}) provided a unified managed memory that
is accessible to both the CPU and GPU, and thus avoids the need of
explicitly programming the migration of data from the CPU to the GPU
and back. Moreover, support for object handling, dynamic allocation
and disposal of memory is rapidly growing. On the software side, many
projects develop high-level programming models for GP-GPU, such as
OpenAcc for C/C++, Copperhead for Python, the NOVA statically-typed
functional language, and X10 heterogeneous compiler. 
This field is clearly not yet mature, but the integration of the GPU
concurrency model with high-level programming is just a matter of
time.

\begin{example}[The evolution of shared memory abstractions]
We can trace an evolutionary lineage also locally to shared memory
abstractions. The original Java threading model provides a class
\sverb{Thread} whose instances are associated to (possibly
dynamically) created JVM threads. To spawn a new thread the programmer
must create a new object of type \sverb{Thread} that encapsulates the
code to be executed in parallel, and then call the method
\sverb{start()} on such an object. However, requesting and obtaining
the exclusive control of 
a JVM thread is costly in terms of performance. Novel solutions
enforce a clear distinction between \emph{logical threads}, that is,
the activities to be concurrently executed, a.k.a. tasks, and the
\emph{executors}, which are pools of (JVM) thread workers. In
particular, the link between the task to be executed and the thread
actually executing it is taken away from the programmer's control and
it is devolved to an efficient work-stealing scheduling algorithm
implemented by the runtime. 
The difference between the lightweight threads spawned in the program
and the pool of executors available in the runtime is particularly
marked in the X10 language. In this language, a block of code to be
executed in parallel is simply defined by the statement
\sverb{async\{...code...\}}. Dually, the \sverb{finish\{...\}}
statement instructs the control to wait for the termination of all
the concurrent code that might have been asynchronously spawned within
the \sverb{finish} block. In other terms, X10 concurrency primitives
provide a very simple (but effective) fork/join model that abstracts
away the management of thread workers. Similar (but less
straightforward) solutions are available in recent Java releases,
together with a number of classes that allow the programmer to
customize the pool of runtime executors. 

To conclude, we observe the evolution of another distinctive feature
of the Java threading model: the use of locks and conditions. As we
discussed above, correctly using primitives like \sverb{synchronized,
  wait(), notify()} is very difficult. Hence Java concurrency library
encapsulates a correct usage of these primitives into ready-to-be-used
higher level abstractions such as atomic values, barriers, 
synchronized data structures. However, X10 completely dismissed Java's
low level, error-prone building blocks, in favor of higher-level
primitives such as 
\sverb{atomic\{...\}} and \sverb{when(condition)\{...\}}, directly
taken from Software Transactional Memories.
\end{example}

\section{The post-concurrency era}
\label{sec:postConc}

\begin{figure}[th]
\begin{center}
\includegraphics[width=12cm]{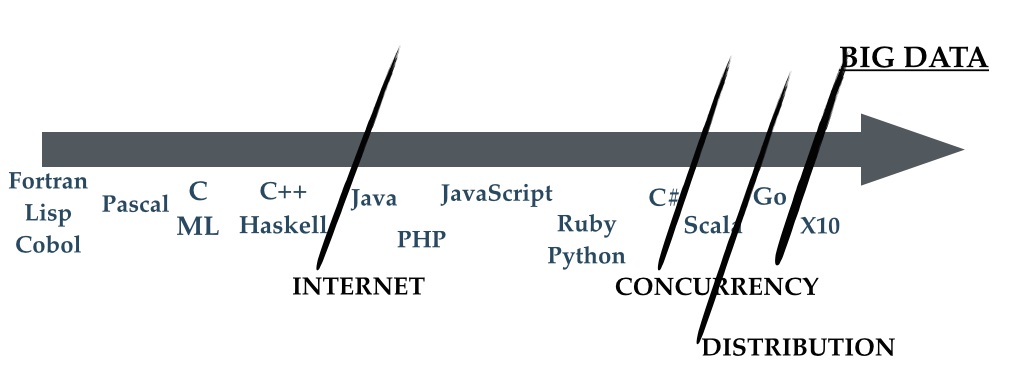}
\vspace*{-0.3cm}
\caption{Complete timeline}
\label{fig:timelineII}
\end{center}
\end{figure}

In this section we complete the overview of the programming languages
timeline that we started in Section~\ref{sec:quest}.
In particular, we discuss a couple of more recent driving forces that
are powering other significant changes in mainstream programming, as
illustrated in Figure~\ref{fig:timelineII}.

\subsection{The impact of Clouds in programming languages}

Besides multicores and GPUs, which drive concurrent programming,
another important achievement of modern technology is cloud computing,
which is acting as a catalyst for \emph{distributed programming}. 
Also in this case, distributed systems are well-established, together
with a number of successful programming solutions like sockets, RPCs
(Remote Procedure Call), RMI (Remote Method Invocation), the
Grid, SOA (Service Oriented Architecture). However, the key change
here is that the popularity of cloud resources shifted the business
applications model from a centralized service provider accessed in
client-server model, 
to the so-called Software As A Service (SaaS) model, where
applications are deployed on heterogeneous platforms, from mobile
devices up to cloud based clusters, running thousands of multicore
processors. Such a model requires ad hoc solutions to productively
deal with scalability issues, hardware heterogeneity, fault tolerance,
security and privacy. These requirements demand new architectural and
system-level solutions, but also well-suited programming models can
clearly offer a strong support.

In this context, \emph{Reactive Programming} is attracting a growing
appeal. The Reactive Manifesto~\cite{ReactManifesto}
precisely distills the distinctive features of reactive applications:
they must be ready to react to events, react to load, react to
failures, and react to users.  
\begin{description}
\item[Reactivity to events] entails an event-driven programming model
  that, instead of issuing a command that \emph{asks for a change},
  reacts to an event that indicates that something \emph{has changed}.
  Such a model endorses all kinds of asynchronous operations: non
  blocking operations and decoupling event generation from event
  processing result in higher performance and scalability. As an
  example, \emph{futures} are an asynchronous concurrency abstraction
  that has come to light in many recent mainstream languages.
\item[Reactivity to load] means the ability to scale up/down to deal
  with addition/removal of CPUs on a node, and to scale out/in to deal
  with addition/removal of server nodes. Scalability, or elasticity,
  requires loose coupling between component behaviour and its location
  (i.e., location transparency). It also requires to minimize shared
  mutable state and explicitly focuses on components communication.
\item[Reactivity to failures] asks for programming styles that enforce
  application resilience, in order to quickly recover from software
  failures, hardware failures, and communication failures. For
  instance, minimization of interdipendencies between components,
  encapsulation and hierarchic supervision lead to software components
  that are better isolated and monitored.
\item[Reactivity to user interaction] addresses application
  responsiveness. This aspect champions asynchronous and event-driven
  approaches and  technologies that push data towards consumers when
  available rather than polling, such as push-servers~\cite{WikiPush}
  or WebSockets~\cite{WikiWebSocket} 
  which push events to browser and mobile applications.
\end{description}

This scenario will act as the environment operating a selection over 
the features of actual programming languages. Many of the features
described in the previous sections clearly have to do with some of the
items above, but cloud computing and reactive programming bring about
a new shuffle of old issues and new problems. Hence new ``language
mutations'' will appear to adapt to these new requirements.

\subsection{The Big Data era}

By reaching the end of our timeline, we find the current challenge we
are facing in computer science: dealing with the huge amount of data
that are collected by smart devices and pervasive computing. 
Big data applications require high-performance and data-parallel
processing at a greater order of magnitude. Interestingly, reactive
programming can be convenient here with its view of data in terms of
streams of data rather than a collection/warehouse of data. However, 
analytic computations on big data are proving to be the killer
application for \emph{High Performance Computing} (HPC), that is, the
programming model designed for scale-out computation on
massively parallel hardwdare. This paradigm is targeted not only to
cloud infrastructures, but also to high-performance computing on
supercomputers with massive numbers of processors. To exemplify, let's
consider the X10 programming language, an open-source language
developed at IBM Research~\cite{X10}, whose design recombines earlier
programming abstraction into a new mix in order to fit the HPC
requirements. X10 is designed around the 
\emph{place} abstraction, which represents a virtual computational
node that can be mapped onto a computer node in the cluster, or onto a
processor or a core in a supercomputer. According to the HPC model, a
single program runs on a collection of places, it can create
global data structures spanning multiple places, and it can spawn
tasks at remote places and detect their termination. 
At the same time, a significant design feature is
that X10 is an object-oriented, high-level programming language: its
syntax is reminiscent of Java-like and Scala-like abstractions, and a
powerful and expressive type system enforces static checks and
promotes the programming-by-contracts methodology in terms of
constrained dependent types. 

Notice that tackling high performance computing by
means of high-level programming abstractions enhances productivity and
integration with mainstream programming. However, it also implies that
considerable efforts must be put in place at the runtime environment
level: raising the abstraction level of the source program must indeed
be balanced out by an efficient executable with significant
performance on heterogeneous hardware. Indeed, in the case of X10, the
source code compiles either to Java code, or to C++ code or also to
CUDA code. Moreover, the resilient runtime supports executions that
are tolerant of place failures. X10 is then well-suited to write code
running on 100 to 10.000 multicore nodes, that is up to 50 millions of
cores. 

As far as concurrency is concerned, high performance computing
requires specific abstractions that allow the program to capture the
logic of big data applications without explicitly dealing with
distribution and parallelism issues. Once again, the quest for
good abstractions that allow both a more declarative code and an
efficient implementation is crucial also in HPC, where the gap between
the logic of the application and the execution infrastructure is
particularly large. A couple of so-called big data application
frameworks are proving to be successful: the \emph{map-reduce} model
and the \emph{bulk synchronous parallel} model. The map-reduce model,
implemented for instance by Google's MapReduce, Apache Hadoop and
Apache Spark, is a model inspired by map and reduce combinators from
the functional programming, through which many real world tasks can be
expressed (e.g., sorting and searching tasks \cite{MR09,MR11}). 
The Bulk Synchronous Parallel (BSP) model, implemented for instance by
Google’s Pregel [11] and Apache Giraph, has recently gained great
interest because of its ability to process graphs of enormous size,
such as social or location graphs, by deeply exploiting the work
distribution on several machines (e.g.~\cite{BSP94,BSP11}). 
It is worth observing
that these two models could be classified as \emph{concurrency
  patterns}, in that, similarly to classical object-oriented design
patterns, they compose basic (concurrent) abstractions into strategic
solutions. 
This observation illustrates how HPC entails an abstraction level,
hence also a level of explanation,
that is higher than that of concurrent programming, that will lead to
the forthcoming landmark in the timeline.

\section{The role of theory in the evolutionary process} 

The timeline we have described so far lists the catalysts that powered
significant changes in mainstream programming. This list also bears
out the co-evolution of programming languages and hardware
technology, which incidentally is evocative of the co-evolution
of the human language and the brain. 
However, in the history of programming
languages there is another important co-evolving lineage, that
corresponds to the advances of the theoretical research.
It is not worth here to distinguish what theoretical
results affected mainstream programming ad what application solutions
powered theoretical work. Such a distinction would require a precise
historical reconstruction, and it is not essential to acknowledge that
both areas have been mutually influenced, thus co-evolved.

Let us rather recall another quote from Robin Milner's Turing 
lecture~\cite{MilnerTuring} talking
about the semantic ingredients of concurrent computation: 
\begin{quote}
``I believe that the right ideas to explain
concurrent computing will only come from a dialectic between models
from logic and mathematics and a proper distillation of a practical
experience [...] on the one hand, the purity and simplicity
exemplified by the calculus of functions and, on the other hand, some
very concrete ideas about concurrency and interaction suggested by
programming and the realities of communication''. 
\end{quote}

The role of the theory in the evolution of programming languages is
then essential for the \emph{dialectic method} devised by Robin
Milner. In his lecture he aimed at distilling the basic semantic
elements of concurrent interaction, but the same holds also in general
for programming primitives: suitable programming abstractions come
from a dialectic between the experimental tests conducted by practical
programming, and the deep mathematical tests conducted by the
theoretical approach. 
Observe that these two kinds of tests are often conducted at different
times, with practical programming occasionally picking out some
theoretical result tested much earlier in a possibly slightly
different context.

The formal languages studied by the theoreticians are indeed well
suited to test new abstractions and new mix of abstractions in a
concise and expressive model. 
In other terms, they allow for experimentation in a
controlled environment. For instance, notions like asynchrony,
locality, scope extrusion, futures, mobility, security, timing,
probability and many other have been studied by many process calculi
both in isolation and in combination. This kind of research is a
definite contribution to what we called the quest for good
abstractions. In evolutionary terms, we could say that theoretical
research tests and promotes language mutations, not necessarily driven
by the actual environment or the short-term
future.\footnote{Interestingly, even in biology mutations are not 
  always driven by adaptation; see for instance the concept of
  spandrel~\cite{spandrel}.} 

Additionally, when working in a formal framework it is
easier to distinguish the different abstraction levels involved in a
given issue. 
Hence it is easier to first pick out just a single level to develop
specific solutions, and then to study the integration of separate 
approaches into a distinct model, explicitly targeted at the
combination of the different abstraction levels.
A nice example of combination of different abstraction levels are the
works of Abadi, Fournet et Gonthier about the translation of
communication on secure channels into encrypted communication on
public channels~\cite{AbadiFournetGonthierI,AbadiFournetGonthierII}. 
In these works the high level primitives for secure communication in a
pi-calculus-like language are mapped into a lower-level,
spi-calculus-like, language that includes cryptographic primitives.
The correctness theorem for such a translation implies that one can
reason about the security of programs in the high-level language
without mentioning the subtle cryptographic protocols used in their
lower-level implementation.

As a final example of productive mix of abstractions resulted from
the dialectic method described above, we recall the recent process of
integration of functional programming into object-oriented and
concurrent programming languages.
An interesting way of looking at such a process
is observing that this integration is actually fostered by
understanding the notion of \emph{function} as an abstraction that
represents a \emph{behavior}, which can be passed around and composed.
Moreover, while imperative programming involves thinking in terms of
\emph{time}, functional programming rather focuses on \emph{space},
where basic functions/behaviors are composed by need as building
blocks, and the execution advances by transforming immutable values
instead of stepwise modifications of mutable state
(\cite{OderskyTalk}).  
The spatial view of functional programming can smoothly fit 
OOP's ability of structuring software systems. On the other hand,
designing a concurrent system in terms of space rather than time
is easier and it allows one to better deal with the intrinsic
nondeterminism. As a result, after fifty years of functional
programming, the distinctive traits of those languages shine in new
languages essentially because they leverage concurrency. An
evolutionary biologist would call such a functional shift
an example of exaptation~\cite{exaptation}.

\section{Conclusions}

To conclude our excursus 
we point up some final comment. 
First we remark that in the realm of natural languages we
know that writing in a language involves thinking in that
language. The same holds also for programming languages: each language
entails a specific programming style, and we know that \emph{what} is
being said (or coded) is shaped and influenced by \emph{how} is being
said (or coded). And the same also holds for formal models and
theoretical frameworks. It is then important to remind that there is
no best model/language, but there might be a best suited
model/language for a given situation. That is why occasionally
changing the working model/language might be beneficial.

Moreover, modern software systems distinctive of innovative
internet-driven companies such as Google, Facebook, LinkedIn,
are actually written using a mix
of languages, creating a sort of ecosystem of programming languages
that interoperate at different abstraction levels. As a consequence, a
productive mix of models together with interoperable primitives is
vital. 

\medskip
A final comment is devoted to our evolutionary look at the modern
history of programming languages. In this paper we essentially
used the evolutionary talking as a metaphor. However, a thorough
discussion about to what extent Darwin's theory of evolution can
be applied to programming languages would be very insightful.
Indeed, it would be quite interesting to answer questions like
what are language mutations and is there a struggle for life in the
language arena?
Are different concurrency models/abstractions an example of mutations
over which the market (or the marketing strategies) will do its
selection action? 
Will only those languages that are equipped with higher
plasticity either in their design choices or in their marketing
strategies survive? It is not an
accident that Oracle embarked on a deep change of the Java platform,
both in the language, the JVM and the programming style, to leverage
Java8's lambda expressions ~\cite{JavaOne13,JavaOne14}; how else could
we call such a change if not a form of adaptation?




\end{document}